\documentclass[english,aps,pre,twocolumn,showpacs,floatfix]{revtex4-1}
\usepackage[T1]{fontenc}
\usepackage[latin1]{inputenc}
\usepackage{amsmath}
\usepackage{graphicx}
\usepackage{amssymb}
\usepackage{dcolumn} % Align table columns on decimal point
\usepackage{bm} % bold math
\usepackage{color}
\usepackage{cleveref}
\usepackage{subfigure}
\usepackage[none]{hyphenat}

\makeatletter

\usepackage{amsfonts}
\usepackage{epsfig}
\usepackage{dsfont}
\usepackage{babel}

\makeatother

\begin{document}
\title{How visas shape and make visible the geopolitical architecture of the planet}

\author{Meghdad \surname {Saeedian}}
\affiliation{Department of Physics, Shahid Beheshti University, G.C., Evin, Tehran 19839, Iran}

\author{Tayeb \surname {Jamali}}
\affiliation{Department of Physics, Shahid Beheshti University, G.C., Evin, Tehran 19839, Iran}

\author{S. \surname{Vasheghani Farahani}}
\affiliation{Department of Physics, Tafresh University, Tafresh, P.O. Box 39518-79611, Iran}

\author{G. R. \surname {Jafari}}
\affiliation{Department of Physics, Shahid Beheshti University, G.C., Evin, Tehran 19839, Iran}
\affiliation{Center for Network Science and Department of Economics, Central European University, H-1051, Budapest, Hungary}

\author{Marcel \surname{Ausloos}}
\affiliation{GRAPES, rue de la Belle Jardiniere 483, B-4031, Angleur, Belgium}
\affiliation{ Humanities group, Royal Netherlands Academy of Arts and Sciences, Joan Muyskenweg 25, 1096 CJ, Amsterdam, The Netherlands}
\affiliation{School of Management, University of Leicester,  University Road, Leicester, LE1 7RH, United Kingdom}

\date{\today}

\begin{abstract}
The aim of the present study is to provide a picture for geopolitical globalization: the role of all world countries together with their contribution towards globalization is highlighted. In the context of the present study, every country owes its efficiency and therefore its contribution towards structuring the world by the position it holds in a complex global network. The location in which a country is positioned on the network is shown to provide a measure of its "contribution" and "importance". As a matter of fact, the visa status conditions between countries reflect their contribution towards geopolitical  globalization. Based on the visa status of all countries,  community detection reveals the existence of $4+1$ main communities. The community constituted by the developed countries has the highest clustering coefficient equal to $0.9$. In contrast, the community constituted by the old eastern European blocks, the middle eastern countries, and the old Soviet Union has the lowest clustering coefficient approximately equal to $0.65$. PR China is the exceptional case. Thus, the picture of the globe issued in this study contributes towards understanding "how the world works".

\end{abstract}

\maketitle

\section{Introduction}
Today we are all witnessing a new wave of \emph{globalization}. This statement goes with the existence of \emph{entanglement} between the countries. This coexistence is formed by the global policy together with the technological developments of the past few decades. The fact of the matter is that the creation of multinational organizations and alliances surely proves one thing; the countries are entangled to each other with reaches far beyond their own geographic borders. The result to this date is the formation of a more and more sophisticated or even complex international organization~\cite{Peng} and global architecture. This proves the need for further studying the global architecture \cite{Jarzombek,Jarzombek1,Miskiewicz,Miskiewicz1,Miskiewicz2,Karpiarz} based on the international relations and collaborations of individuals, governments, and industrial (or economic) companies~\cite{Namaki1}. The outcome of these collaborations is the flow of ideas and technology, together with goods and human resources throughout the world, which, as a byproduct, helps  to diminish the influence of geographical borders on human affairs. One of the main things that restrains globalization is \emph{geopolitical architecture}, which describes the ways in which organizations deal with flows and borders \cite{Dodds}.

Historically, the world has been somehow geopolitically structured~\cite{Kenneth,William,Veugelers}. The structure has always owed its existence to the dominant will and enthusiasm of that era~\cite{Durant}. As such, local rivalries of two, sometimes a few more, countries shaped the local structure of the globe; see the Trojan War ca. 1100 BC, the rivalry between Persia and Greece beginning in ca. 490 BC, the France-England rivalry leading to the Hundred Years' War stretching from 1337 to 1453, etc. Physical structures, like Hadrian's Wall or the Great Wall of China, which were built to limit intrusion and to define the domain of the ruling power, are materializations of geopolitical structures of that era. Another manifestation of such structures are the travel papers which were issued in medieval and ancient times after the applicant had paid the appropriate amount at the pay toll located at the border~\cite{Jean}. Although, these structures have been proved to be rooted in, or rather explained through, politics, religion, economics, and military issues, they owe their existence and survival to \emph{power}. Achieving power \cite{French} is an objective not only for living beings but also for organizations, societies, and countries. Although in today's world the plans for achieving triumph in various aspects have conceptually changed compared with the past \cite{Kenneth,Deutsch,Baylis,Shirazi}, achieving power as done yesterday still needs organized action~\cite{CrozierFriedberg}.

In ancient times, the empire with a greater army (in the sense of number of warriors) was considered mightier by everyone, even though the battle field topology or battle strategy could influence the outcome of the battle~\cite{Fegley,Hipshon}. This thought remained mostly correct till the 13th century when firearms were invented \cite{Brit}. An army equipped with firearms and heavy artillery proved more effective compared with a larger army only equipped with sharp swords and swift horses; the battle of Chaldoran between the Persian and Ottoman empire is a perfect example~\cite{Camb}. The never ending development of weaponry handed the power to the countries with more sophisticated weapons~\cite{Kenneth,Deutsch}. Based on this argument, one would think that nowadays, a powerful country is a country that possesses more war planes and ballistic missiles. Maybe this was true up to a couple of decades ago. But does this statement work today? To answer this question, one must first understand the concept of power in modern world and more importantly its features and structures. Only then, one could confidently understand why some countries with small armies or even no army are very effective in the balance of power, when previously, such countries were nowhere near to be considered as possible decision makers. Today, neglecting countries without "active armies" (e.g., Japan and Germany) is absolutely irrational, - since they also play an effective role in the architecture of the nations, on various geo-socio-political levels. What do these countries have that  makes them so influential? The answer lies in the concept of the invisible power of \textit{community} ~\cite{Basu,Oleinik}.

In the present study, we focus on the term \emph{community}, highlight its features, and make visible the role of communities in the geopolitical architecture of the world. A select candidate for studying these communities and their role is the visa status between countries which is less dependent on economic considerations but more tied to political and social considerations. How  this is possible and how  a visa constraint can model a community  are illustrated in the next section. However, first of all, it is instructive to recall ideas leading towards the creation of visa.

Demand for travel papers (or as said today, passport) when crossing country borders dates back to ancient times. Obtaining visa or clearance for inter country travels was unnecessary before the First World War, at least in Western Europe. During World War I, the visa came as a stamp in the passport which allowed entrance in a country. By the end of the War, the concept of visa had already obtained its significance. As a matter of fact, the widespread implementation of visa among European countries was triggered by the aftermath of the First World War due to security reasons~\cite{Stearns}. Today, not only security but also geopolitical, economical, technological, and scientific issues oblige the use of visa. ~\cite{Lepgold,Stern}.

However, due to the race for achieving a higher political, economical, cultural, and scientific position among countries, the process of visa has either been eased or waived~\cite{Jaroszewicz,Stent}. However, a visa free status between two countries or more (recall the Shengen agreement between 26 countries \cite{Sheng}) can be accepted as a realistic proof of "positive interactions". These interactions provide the backbone of communities. To identify these hidden communities, we analyze a network in which every two nation that have reciprocally waived visa between them are linked.

In this work, the complex network approach is implemented to sketch the architecture of the world. By studying this architecture, a better appreciation of the various relations between countries can be obtained. This understanding would simply enable one to observe the consequences of cutting relations with other countries, i.e. in some sense, cutting relations with a country which belongs to a community, would suggest that some collective behaviour (as part of the networks evolution) against that country is likely to occur. Through this study, we make visible the most influential communities. Using the concept of clustering coefficient, we assign a value to each of the communities as a measure of likelihood of two positively related countries to be in positive relations with the third. This measure can be interpreted as a degree of globalization of a community.

\section{A network named visa}
The network studied here is constituted by countries, where the type of relation between countries marks their type of links. If the citizens of country ``A'' need a visa to enter country ``B'' and vice versa no link would exist between node A and B. But if country A waives a visa  requirement for the citizens of country B, an arrowed link from node B to A is drawn and vice versa.

\begin{figure}[t!]
\centering
\includegraphics[width=1.0\linewidth]{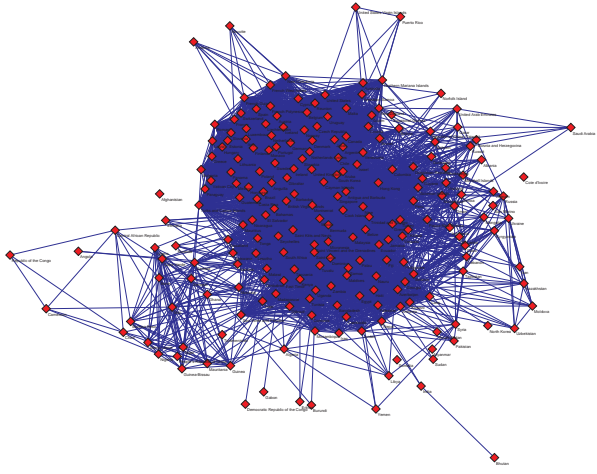}
\caption{(color online) The reciprocal visa free network. Each link between two nodes (countries) is an indication of a reciprocal visa free statues between them. This network contains 222 nodes and 5582 links. The nodes are labeled by the names of countries which could be better seen by zooming in.}
\label{fig: 1}
\end{figure}

Presently 222 countries exist in the world~\cite{Doyou}. Therefore our network is constructed by the existing or non-existing directed links between every 222 country with the other 221 countries due to their visa free status. Although, as it is well known, there are many nations for which their citizens need a visa to enter many countries, still many two way directed links between countries exist in addition to one way directed links, e.g., USA and several EU countries. The resulting network which we choose hereafter to call the {\it visa free network} consists of 21383 directed links which 10219 of them are one way \cite{Doyou}. In Fig.~\ref{fig: 1} the links between the nodes is an indication of a reciprocal visa free status between two countries. In the case of a one sided visa free situation between two countries, no link is drawn. Also if the citizens of both countries need a visa to enter the other country, again no link is drawn. In our opinion, good relations between two countries requires that both sides waive visa for the citizens of the other. That is why only in such a situation the two countries would be linked.

\begin{figure}[t!]
        \centering
        \subfigure[]{
            \centering
            \includegraphics[width=0.215\textwidth, height=0.17\textwidth]
            {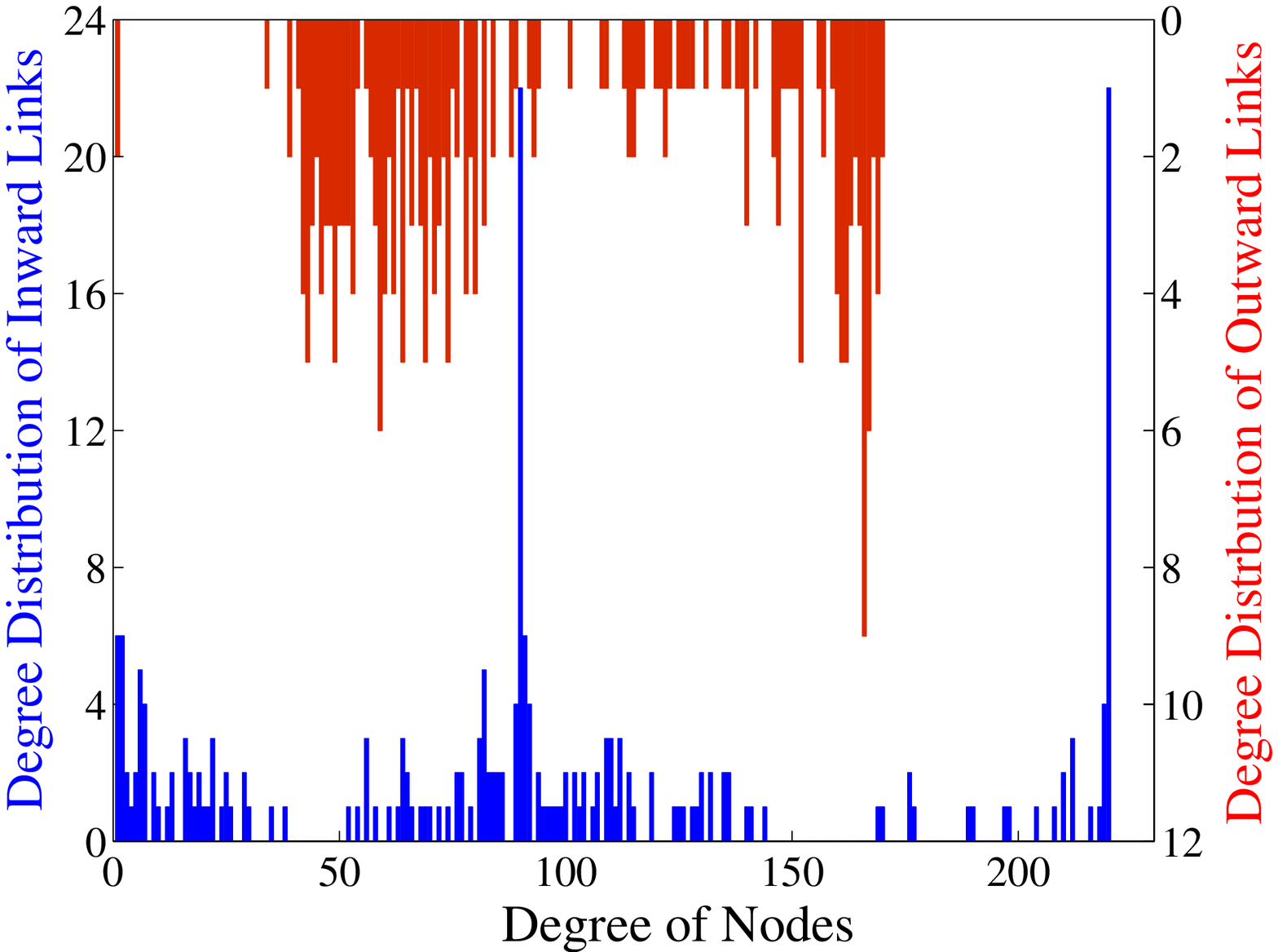}
            \label{fig: in- and out-degree distribution}
        }
        ~ %add desired spacing between images, e. g. ~, \quad, \qquad, \hfill etc.
          %(or a blank line to force the subfigure onto a new line)
        \subfigure[]{
            \includegraphics[width=0.215\textwidth, height=0.17\textwidth]
            {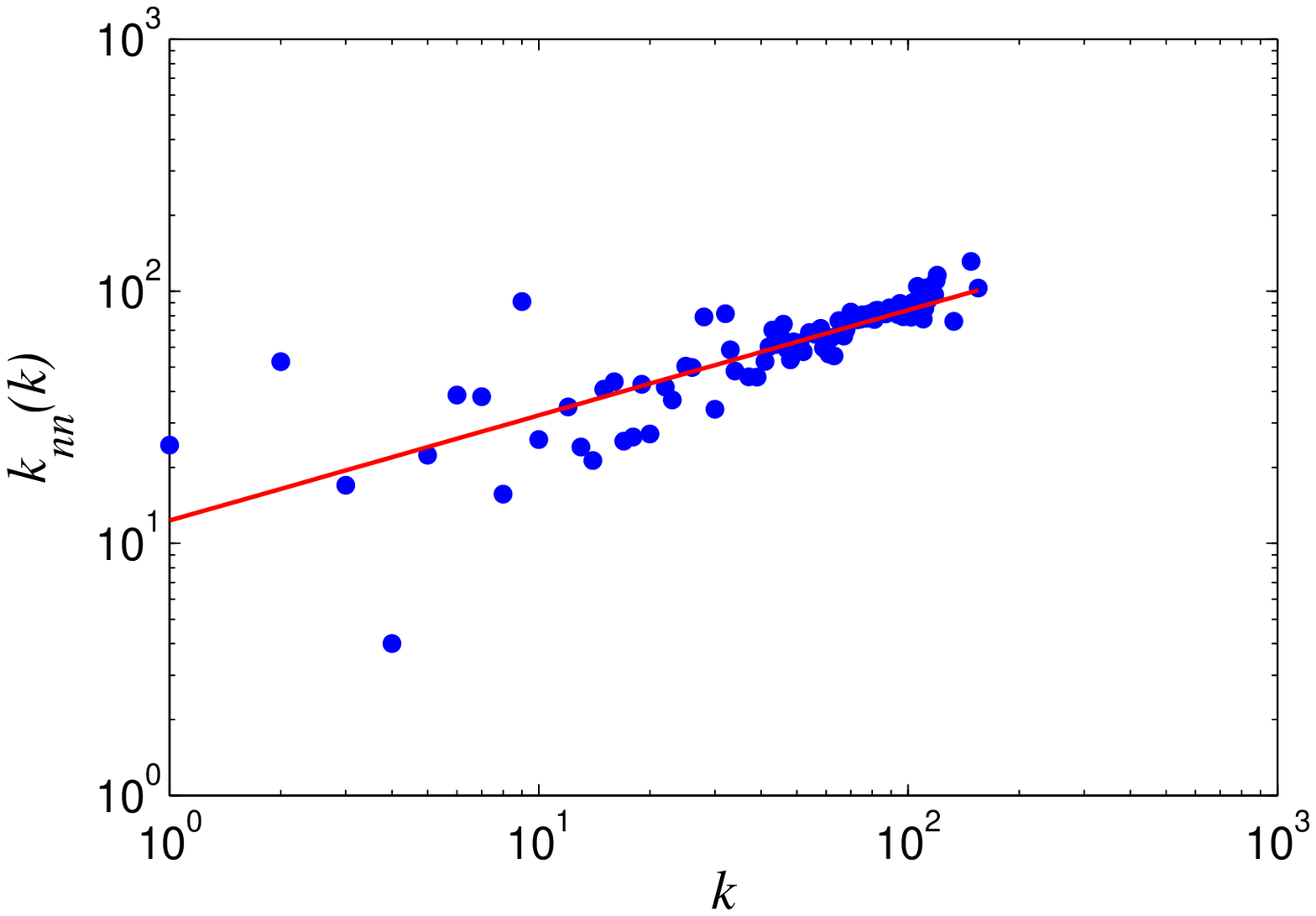}
            \label{fig: degree correlation function}
        }
        \caption{(color online) Left panel; the in-degree (blue bars) and out-degree (red bars) distributions. For the in-degree distribution there are three distinguished regions characterized by  three peaks. The left region consists of countries with a few inward links: this means that almost every citizen of the world needs visa for entering them. This clearly indicates that these countries possess a high level of security. The middle region is around the middle peak which is mainly constituted by European countries. The third region  on the extreme right belongs to the countries with many inward links, meaning that almost every citizen of the world could enter them without visa. This region mainly consists of countries with a strong tourism industry. For the out-degree, unlike the in-degree, distribution there is only one region which is localised around the middle.
        Right panel; the degree correlation function $k_{nn}(k)$ for the visa free network. The power law behaviour proportional to $k^{-\nu}$ is observed. The fit is due to a value of $\nu$ equal to $0.42$. The positive sign for $\nu$ indicates that the network is assortative.}
        \label{fig: 2}
\end{figure}

One of the main characteristics of the directed complex network is its in- and out-degree distributions \cite{Albert,YYY1,YYY2}. The in- and out-degree distributions are shown in the left hand panel of Fig.~\ref{fig: 2} (blue and red bars respectively). It can readily be noticed that there are three (blue) regions around three peaks. The first region is around the left peak which is located at the origin $(k_{in} \sim 0)$, indicating  that countries "near the origin" have a small in-degree (inward links) value. This means that almost every citizen of the world needs a visa in order to enter these countries. 
The second region is around the middle peak which is located at $k_{in} \sim 90$. The third region is around the farthest right peak which is located at $k_{in} \sim 220$:  for these,  $almost$ every citizen of the world could enter the countries near the right peak $without$ visa.

If one considers more closely the countries comprising each of the three sets, it can be observed  that the first region mainly consists of countries with a high level of security; the second region  mainly consists of European countries, and the third region mainly consists of countries with a strong tourism industry.

The red bars in Fig.~\ref{fig: 2} show the out-degree distribution function. It can be readily noticed that the out-degree bars are much more localized compared with the in-degree bars: the central peaks of such distributions occur (i) at $ k_{out} \sim 55$ and (ii) $k_{out} \sim 160$ respectively. This is showing  that the out-degree distribution is mainly peaked between the $k_{in}$ values. This means that at places on the left and the right of the in-degree distribution, the out-degree distribution has a zero value.

To comply with the aims of the present study which is studying the communities created by the friendships between nations, we must look at countries that have reciprocally waived the visa requirement between them. The reason for this is that when both sides (countries) reciprocally wave the visa requirement between them, their relations would be more robust ("friendly"); any result deduced from their relations should appear to be much more conclusive as compared with the case when only one of the sides is visa free for the other.

For the network being so characterized, we need to take a further step and measure the degree correlations of the visa network. The degree correlations represent the way in which the degree of a node is related to the degrees of its neighbors. Studies show that most social networks have high assortative degree correlations \cite{Newman2002}. "Assortative"  means that nodes are preferentially   connected with nodes being peers (and equal) to themselves. In other words, nodes with high degrees are preferentially connected with other nodes with considerable degrees, while nodes with low degrees are preferentially connected with other low degree nodes. The term assortativity, in the context of the present study, means that countries with high degrees (low degrees) have a tendency to be connected with other countries with high degrees (low degrees). In contrast to an assortative network, a disassortative network is a network for which  its countries with high degrees are less connected with each other, but instead are mostly connected to countries with low degrees.

\begin{figure}[t!]
\centerline{\includegraphics[width=1.0\linewidth]{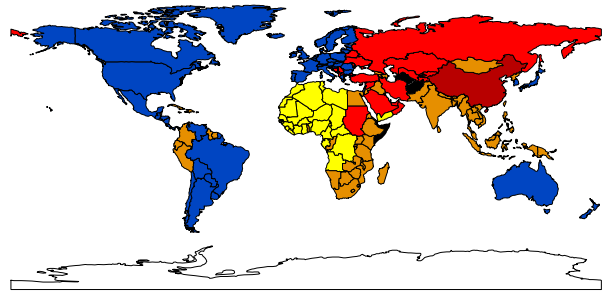}}
\caption{(color online) The four plus one communities constituting the globe. Countries in the same community are in the same color; the blue community consists of developed countries, the red community mainly consists of European eastern blocks plus some middle eastern countries, the yellow community consists of African countries, and the beige community mainly consists of south eastern African and Asian countries. This picture clearly shows the neighbouring effects, in a sense that countries of the same community are located close to each other. }
\label{fig: communities in the world map}
\end{figure}

In order to examine whether the visa network is assortative or not,  one calculates the degree correlation function $k_{nn}(k)$ which is defined as the average degree of neighbors of a node with degree $k$ \cite{Newman2002}. Mathematically speaking, the degree correlation function is given by
\begin{equation}
\label{eq: degree correlation function}
k_{nn}(k)=\sum_{k^\prime} k^\prime P(k^\prime|k),
\end{equation}
where the conditional probability $P(k^\prime|k)$ is the probability that a link of a node with degree $k$ points to a node with degree $k^\prime$. In Fig.~\ref{fig: 2} the degree correlation function $k_{nn}(k)$ for the visa free network exhibits a power-law behavior as $k^{-\nu}$, with $\nu\approx 0.42\pm 0.07$. Since the slope is positive, the visa network  can be claimed to be assortative. Having understood the assortative level of the the visa network,  one can further comment on the communities. In order to perform a "community detection process" we follow \cite{Newmac}, and obtain the colored world map, Fig. \ref{fig: communities in the world map}.  Note that the selection criterion for the colors or in other words the criterion that puts certain countries in the same division is based on a simple matrix method; where the network is continuously divided until the dividing of a sub-graph would not increase the modularity of the network; see \cite{Newmac} for details.  By mere looking at Fig. \ref{fig: communities in the world map} it
 could be readily confirmed ("due to" the  coloring Blue, Red, Yellow, Beige) that there exist four main communities plus a community consisting of one country which is PR China \cite{Wad}. However China and the countries in black do not belong to any of the four main communities. We have manually changed the color of China from black to brown in order to emphasize that PR China is not only a highly populated country but has also a high GDP growth rate, different from the few  "black countries". This makes China to be considered as a truly "emerging" country, but with a considerable influence in the world. Since, in the present study (recall the introduction), we are tying the communities to their "power",  PR China itself proves adequate to be considered as a specific community indeed.

\begin{figure*}[t!]
\centering
    \subfigure[]{
        \centering
        \includegraphics[width=0.4\textwidth]{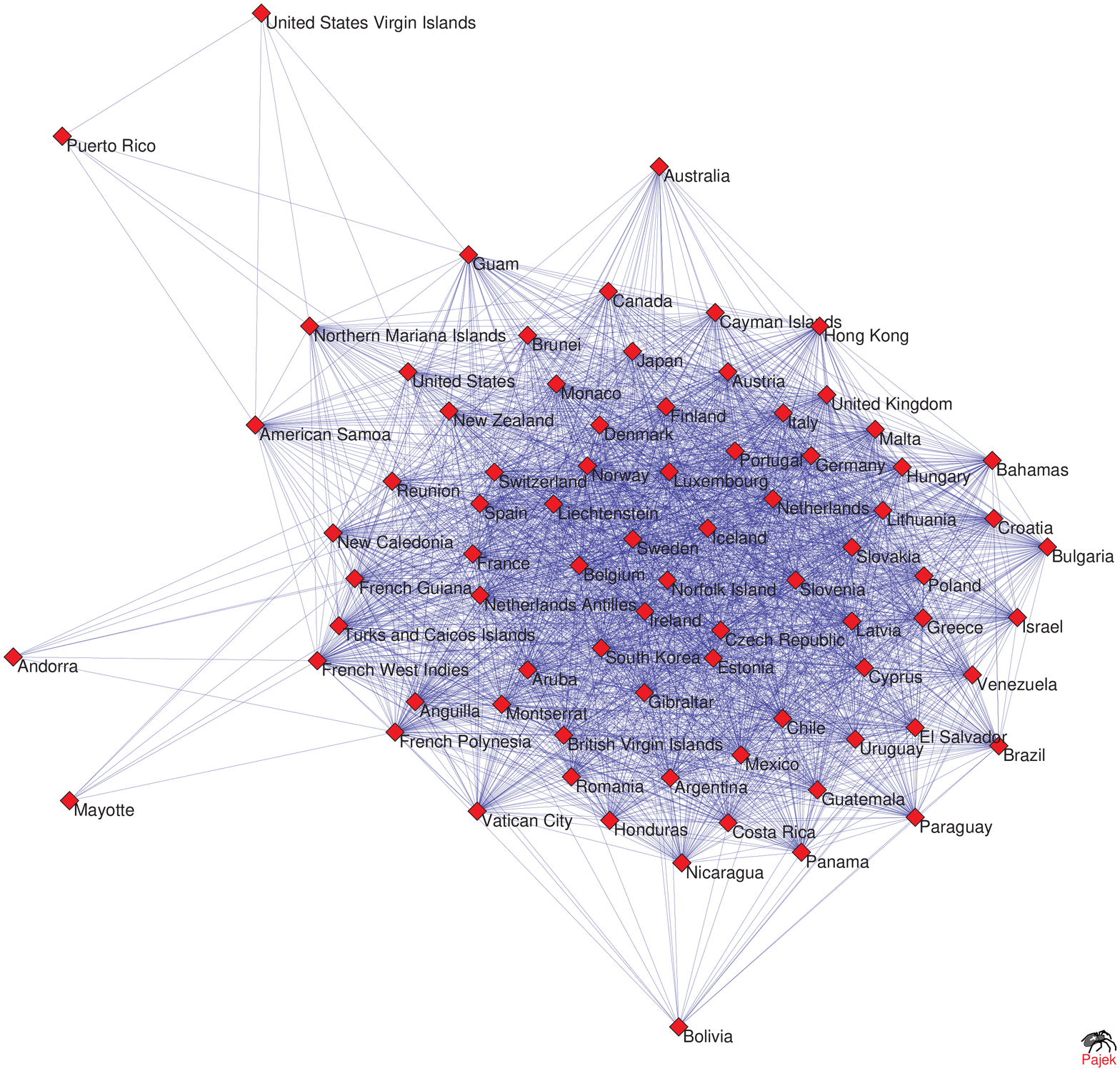}
        \label{fig: blue community}
    }
    \quad
    \subfigure[]{
        \centering
        \includegraphics[width=0.4\textwidth]{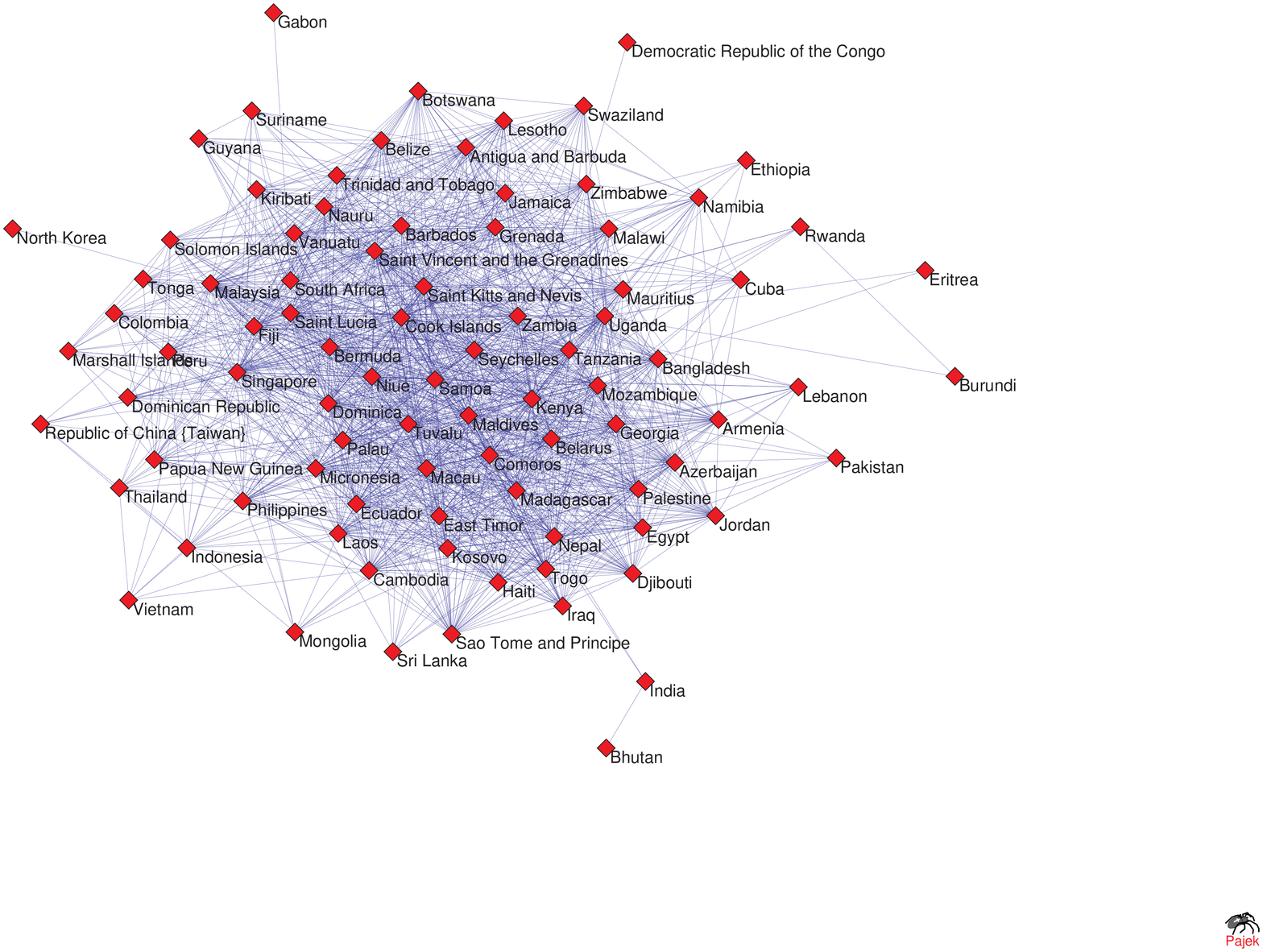}
        \label{fig:subfigure2}
    }
    \subfigure[]{
        \centering
        \includegraphics[width=0.4\textwidth]{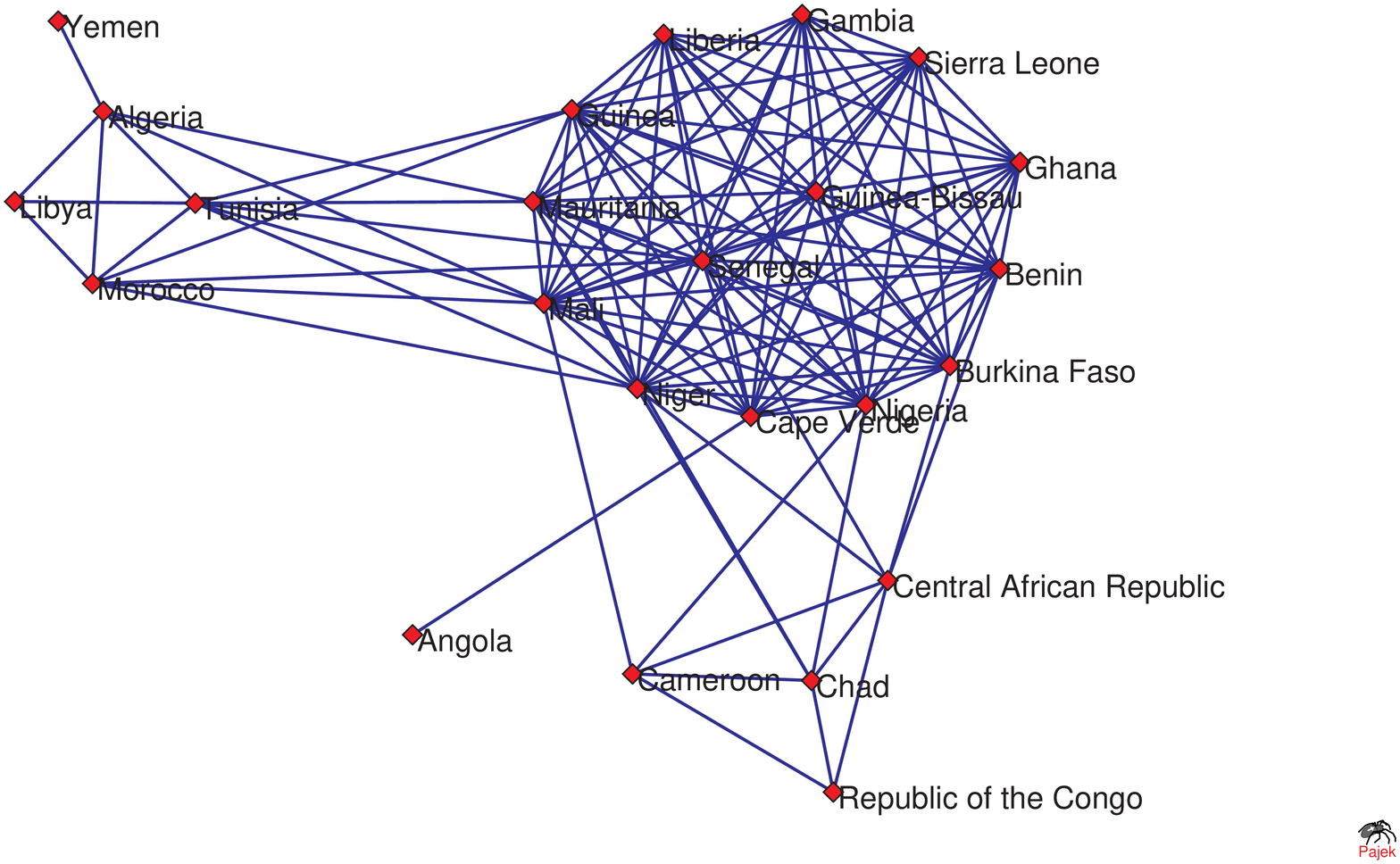}
        \label{fig:subfigure3}
    }
    \quad
    \subfigure[]{
        \centering
        \includegraphics[width=0.4\textwidth]{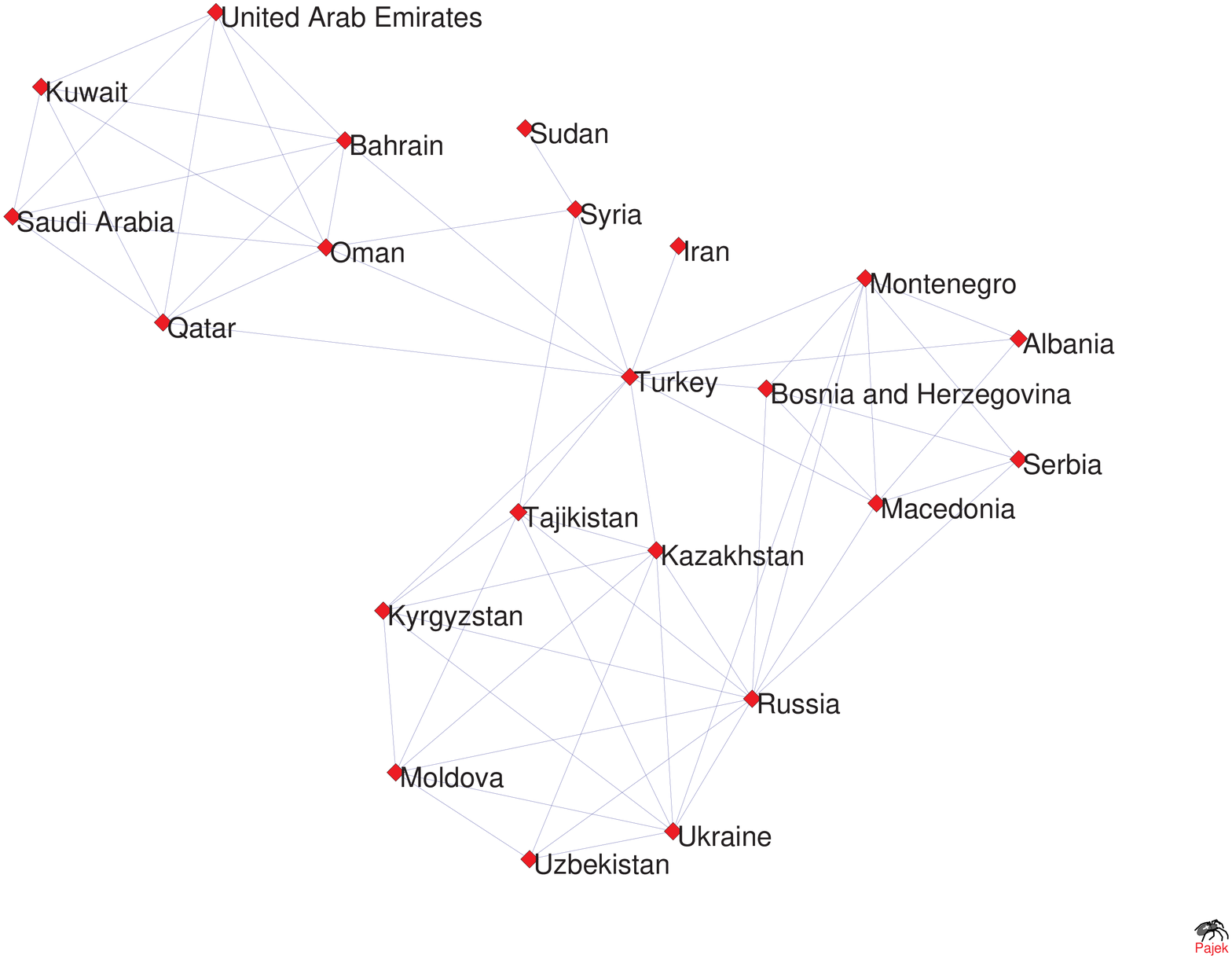}
        \label{fig: red community}
    }
\caption{The community detection provides four main communities in the globe plus China. The top left panel which is for the developed countries (blue countries of Fig. \ref{fig: communities in the world map}) has the greatest clustering coefficient ($0.9$), proving the highest alliance between its members. The bottom right panel has the lowest clustering coefficient ($0.6$) proving the weakest alliance between its members (red countries of Fig. \ref{fig: communities in the world map}). The top right and bottom left panels are for the beige and yellow countries of Fig. \ref{fig: communities in the world map} respectively. }
\label{fig: 4 communities}
\end{figure*}

From Fig. \ref{fig: communities in the world map}, another evidence emerges: the regional effects, somewhat already pointed out in geopolitical analyses of the European Union, through market, civic and  cultural criteria, some time ago \cite{Nicholas}. In the world, countries close to each other also seem mostly to belong to the same community. By glancing over these communities, it is understood that the blue community consists of "developed countries", the red community mainly consists of European eastern blocks plus some middle eastern countries, the yellow community consists of African countries, and the beige community mainly consists of south eastern African and Asian countries.

At this stage, it is worth evaluating the stability of the communities, whence, the clustering coefficient of the network communities \cite{Gligor,Gligor1,WS1}. The strongest clustering goes for the blue community (Fig.~\ref{fig: 4 communities} (a))
with an amount equal to $0.9$, while the weakest clustering goes for the red community (Fig.~\ref{fig: 4 communities} (d)) with a value equal to $\simeq 0.65$. The large clustering number for the developed countries (blue community) is a sign of their alliance together with their trust. In contrast, the small clustering number for the developing countries (red community) could indicate a lower level of trust even among the community members. This may possibly be rooted in memories from the old East European block. By browsing inside the red community, it can be noticed that it is in fact comprised of three sub-communities; the old east European blocks, the countries around the Persian gulf, and countries gained independence from the Soviet union. Interestingly, there is one country that links all of these three sub-communities together, and that is Turkey. In fact, Turkey also connects countries from these three sub-communities geographically  wise. However,  notice that Russia also acts as a link between sub-communities. This means that the betweenness centrality of Turkey and Russia is greater compared to that of other countries.

\section{Conclusions and summary}
In this study, the geopolitical architecture of planet Earth has been visualized. This visualization is based on the visa status between every two country in the world. Visas is a feature of modern life:  a country for which its citizens are allowed to enter more countries without needing a visa automatically saves time for its traveling citizens. But is that all? looking back, there existed tolls between countries or boundaries (walls) between tribes where any suspicious action around the marked or unmarked territorial areas even by an individual was closely observed by the authorities. Now how is this related to power? The answer to this question lies in the conclusions of this study.

Historically, weak tribes and countries had less chances of survival, due to defects in numbers and/or weaponry. Maybe, they could have survived only if a powerful country had backed them. In case of being left alone, some stronger country would have conquered them and had added them to its own territory; recall Caesar conquering Gallia, Alexander of Macedonia going over the Mediterranea, military campaigning through Asia and northeast Africa, creating one of the largest empires of the ancient world, stretching from Greece to Egypt into northwest India and modern-day Pakistan, Cyrus; Genghis Khan, Pizarro, Napoleon Bonaparte, ... But due to the fact that time has changed, such a phenomenon is not experienced anymore, e.g., G.W. Bush is considered as the tenth most largest conqueror \cite{Bush}. This is exactly a feature of modern life which completely defers to what it was two thousand years ago which seemed more to obey the law of the jungle. Nowadays, the reason that modern world rules out the elimination of smaller or weaker countries, is a proof of an "ecological change" in human life. This change owes its existence to the development of mankind leading to globalization. As a matter of fact, it is this globalization that discriminates today from yesterday, and leads to the present planet geopolitical architecture.

Physically, no organization could exist without a structure. As such, the concept of globalization is no exemption, and therefore needs to have a structure. This study has been carried out to focus on this structure. Our working tool for visualising the structure of the world is the concept of visa. By studying the visa status between countries the level of friendship between them is highlighted, which sheds light on the positive interactions they have. In fact it is these interactions that link countries to each other which adds them up to perform communities. Our conclusions are summarized as follows;

- The global network is assortative and therefore its architecture highly depends on the collaboration and mutual interactions of nations.

- The community detection of the global network indicates that the world possesses four plus one main communities, where the two largest communities are in fact the ones with the highest and lowest clustering numbers. The community with the largest clustering number is constituted by the developed countries, while the community with the lowest clustering number is constituted by three sub communities, namely the old eastern European blocks, the middle eastern countries, and the old Soviet union.

- China is a community by itself. This could be due to its rapid growth in economy, industry, and population.

- Regional effects are clearly observed in construction of the communities.

- The connection percentage in each community measures its internal friendship. The highest internal friendship goes for the community of developed countries, while the lowest goes for the community of the old eastern European blocks, the middle eastern countries, and the old Soviet union.

The final word; although most countries are separated by borders, and strictly speaking for citizens of any country their surly exists one or more countries that requires visa for their entrance, but still we have concluded that globalization has been achieved. This means that in order to claim our world as globalized, their is no need for lifting the borders or waiving visa. The existence of hurdles like borders and visas could not oppose globalization. In fact, flowing information, ideas, and goods in the presence of all the hurdles is itself a sign of globalization. Therefore, in the presence of lots of unconnected countries in our planet, globalization could still be enjoyed.

\end{document}